\documentclass[english,aps,nofootinbib,longbibliography,notitlepage,superscriptaddress,prd,preprintnumbers,showkeys]{revtex4-1}
\usepackage{amsmath}
\usepackage{amssymb}
\usepackage{comment}
\usepackage{color}
\usepackage{microtype}
\usepackage{hyperref}
\usepackage{cleveref}
\usepackage{graphicx}

\makeatletter

\@ifundefined{textcolor}{}
{%
 \definecolor{BLACK}{gray}{0}
 \definecolor{WHITE}{gray}{1}
 \definecolor{RED}{rgb}{1,0,0}
 \definecolor{GREEN}{rgb}{0,1,0}
 \definecolor{BLUE}{rgb}{0,0,1}
 \definecolor{CYAN}{cmyk}{1,0,0,0}
 \definecolor{MAGENTA}{cmyk}{0,1,0,0}
 \definecolor{YELLOW}{cmyk}{0,0,1,0}
}

\newcommand{\dd}{\mathrm{d}}
\newcommand{\mn}{{\mu\nu}}

\allowdisplaybreaks

\usepackage{subfigure}

\makeatother

\usepackage{babel}

\makeatletter
\DeclareRobustCommand{\rcite}[1]{%
  \rcite@aux#1,\@nil{#1}%
}
\def\rcite@aux#1,#2\@nil#3{%
  \if\relax#2\relax
    Ref.~\cite{#3}%
  \else
    Refs.~\cite{#3}%
  \fi
}
\makeatother

\begin{document}

\title{Instabilities in tensorial nonlocal gravity}

\author{Henrik Nersisyan}
\email{h.nersisyan@thphys.uni-heidelberg.de}
\affiliation{Institut f\"{u}r Theoretische Physik, Ruprecht-Karls-Universit\"{a}t Heidelberg,
Philosophenweg 16, 69120 Heidelberg, Germany}

\author{Yashar Akrami}
\email{akrami@lorentz.leidenuniv.nl}
\affiliation{Institut f\"{u}r Theoretische Physik, Ruprecht-Karls-Universit\"{a}t Heidelberg,
Philosophenweg 16, 69120 Heidelberg, Germany}
\affiliation{Institute Lorentz, Leiden University, PO Box 9506, Leiden 2300 RA, The Netherlands}

\author{Luca Amendola}
\email{l.amendola@thphys.uni-heidelberg.de}
\affiliation{Institut f\"{u}r Theoretische Physik, Ruprecht-Karls-Universit\"{a}t Heidelberg,
Philosophenweg 16, 69120 Heidelberg, Germany}

\author{Tomi S. Koivisto}
\email{tomi.koivisto@nordita.org}
\preprint{NORDITA-2016-102}
\affiliation{Nordita, KTH Royal Institute of Technology and Stockholm University,
Roslagstullsbacken 23, 10691 Stockholm, Sweden}

\author{Javier Rubio}
\email{j.rubio@thphys.uni-heidelberg.de}
\affiliation{Institut f\"{u}r Theoretische Physik, Ruprecht-Karls-Universit\"{a}t Heidelberg,
Philosophenweg 16, 69120 Heidelberg, Germany}

\author{Adam R. Solomon}
\email{adamsol@physics.upenn.edu}
\affiliation{Center for Particle Cosmology, Department of Physics and Astronomy,
		University of Pennsylvania, Philadelphia, Pennsylvania 19104, USA}

\begin{abstract}
We discuss the cosmological implications of nonlocal modifications of general relativity containing tensorial 
structures. Assuming the presence of standard radiation- and matter-dominated eras, we show that, except in very particular cases, the nonlocal terms contribute a rapidly growing energy density. These models therefore generically do not have a stable cosmological evolution.
\end{abstract}

\keywords{modified gravity, nonlocal gravity, dark energy, background cosmology}

\maketitle

\section{Introduction}\label{sec:intro}
Most extensions of general relativity are manifestly local. \textit{A priori}, however, we need not impose this restriction. Like general relativity itself, most proposed theories of modified gravity are nonrenormalizable, which is often a sign of new physics at high energies. From a local high-energy theory, nonlocalities often appear in the effective theory describing low-energy physics. For example, nonlocalities appear generically when massless or light degrees of freedom are integrated out of a local fundamental theory \cite{Barvinsky:1987uw,Barvinsky:1990up,Barvinsky:1990uq,Barvinsky:1995jv}. 
 
Nonlocal modifications of general relativity constructed out of inverse differential operators give rise to infrared effects 
that become relevant at large temporal and spatial scales. The consequences of these nonlocalities are far-reaching and could provide a dynamical explanation for dark energy. Numerous examples of this line of thinking can be found in the 
literature \cite{Deser:2007jk,Deser:2013uya,Woodard:2014iga,ArkaniHamed:2002fu,Dvali:2006su,Dvali:2007kt,Maggiore:2013mea}.
Most of the existing nonlocal gravity models are purely phenomenological and are constructed out of nonlocal 
operators involving the Ricci scalar only for reasons of simplicity \cite{Maggiore:2014sia,Foffa:2013vma,Dirian:2014ara,Nersisyan:2016hjh}. It is still an open question whether we should expect these particular nonlocal structures, as opposed to something more complicated, to arise in the low-energy limit of fundamental theories. Tensorial extensions involving elements such as the Ricci or the Riemann tensors should not be \textit{a priori} excluded.

Adding nonlocal interactions can also improve some of general relativity's more undesirable properties, and these seem to specifically require tensorial nonlocalities. For example, in order to alleviate the ultraviolet divergences of general relativity, one has to modify the graviton propagator, which requires a tensorial term in the action \cite{Biswas:2011ar}. Considerable recent progress has been made in ghost-free ultraviolet nonlocal gravity \cite{Talaganis:2016ovm,Modesto:2016max}. Furthermore, nonlocal modifications of general relativity could degravitate a large cosmological constant, providing an appealing solution to the problem of why a large vacuum energy does not gravitate~\cite{ArkaniHamed:2002fu}. For the purposes of degravitation, it is likely insufficient to rely on scalar degrees of freedom introduced via nonlocal scalar curvature terms. Tensorial nonlocalities, by contrast, could help implement a consistent degravitation mechanism, as is the case in the framework of massive gravity where nonlocalities modify the tensor propagator~\cite{Dvali:2007kt,deRham:2007rw}.

The cosmological consequences of tensorial nonlocalities involving inverse powers of the d'Alembertian operator 
were considered in \rcite{Ferreira:2013tqn,Cusin:2015rex,Tsamis:2014hra}. Tensor nonlocalities in these models were shown to 
contain rapidly growing modes, leading to instabilities in the background expansion.\footnote{Note that these instabilities are not directly related to unitarity violation. This can be seen by considering a nonlocal term of the form $G_{\alpha\beta}(m^{2}/\Box^{2})R^{\alpha\beta}$, which results in a massive graviton propagator of a unitary form (cf. \rcite{Dvali:2006su}, where this model corresponds to $\alpha=0$ and is shown to be unitary.). However, as the action contains, in addition to scalar terms, the tensorial term $R^{\alpha\beta}(m^{2}/\Box^{2})R_{\alpha\beta}$, it will be cosmologically unstable, as can be seen from Ref.~\cite{Ferreira:2013tqn} and from the following.} Note, however, that the 
inverse d'Alembertian operators considered in these references are certainly not the most general possibility 
that can be implemented at each order in curvature. It is possible that other well-motivated 
differential operators might lead to a somewhat different evolution that is consistent with observations. 

In this work we extend the analysis of \rcite{Ferreira:2013tqn,Cusin:2015rex} to general nonlocal tensorial 
actions at quadratic order in the curvature invariants and investigate whether these modifications are phenomenologically viable. 
This paper is organized as follows. In Sec.~\ref{sec:model}, we introduce our tensorial nonlocal model. The 
cosmological consequences of this model during radiation (RD) and matter domination (MD) are discussed in Sec.~\ref{sec:cosmo}. Finally, the conclusions are presented in Sec.~\ref{sec:conclusions}.

\section{The $R_{\alpha\beta}\triangle^{-1}R^{\alpha\beta}$ model}\label{sec:model}

Consider the most general action quadratic in the curvature invariants \cite{Biswas:2011ar} for some differential operator $\triangle$,
\begin{equation}
S=\frac{M_{\rm Pl}^2}{2}\int\text{d}^{4}x\sqrt{-g}\left(-R+Rf(\triangle)R+
R^{\alpha\beta}g(\triangle)R_{\alpha\beta}+R^{\mu\nu\alpha\beta}h(\triangle)
R_{\mu\nu\alpha\beta}\right)+\int\text{d}^{4}x\sqrt{-g}\mathfrak{\mathit{\mathcal{L}_{m}}}\,,\label{eq:NLAction}
\end{equation}
where $M_{\rm Pl}\equiv (8\pi G)^{-1/2}$ is the reduced Planck mass and $\mathfrak{\mathit{\mathcal{L}_{m}}}$ is the matter Lagrangian minimally coupled to gravity.
Different nonlocal theories are characterized by different choices of the operator $\triangle$ and of the functions
$f$, $g$ and $h$.
In the case $\triangle=\Box$, the above action is the
most general parity-invariant quadratic curvature action; see 
Ref.~\cite{Conroy:2014eja} for derivation of the field equations. We generalize this by allowing for more general differential operators, in particular those with curvature dependence. Note that we would recover the results of 
\rcite{Biswas:2011ar,Conroy:2014eja} for the quadratic truncation of the 
theory.

It is well motivated to consider more general forms for the operator 
$\triangle$; in fact the main rationale for the usual choice $\triangle=\Box$
is just simplicity. For the nonlocally modified theory to be consistent on
suitable backgrounds, one may need to implement a regularization
\cite{Wetterich:1997bz,Barvinsky:2014lja}. For example, \rcite{Barvinsky:2014lja} considered a curvature-dependent regularization
of the form $(\Box+\hat{P})^{-1}$ with\footnote{Here and in the following, $(\mu\nu)$ denotes symmetrization over the indices and $[\mu\nu]$ denotes the antisymmetrization.} 
\begin{equation}
\hat P\equiv P_{\alpha\beta}^{\;\;\;\mu\nu}
    =a R_{(\alpha\;\;\beta)}^{\;\;\,(\mu\;\;\,\nu)}
    +b \big(g_{\alpha\beta}R^{\mu\nu}
    +g^{\mu\nu}R_{\alpha\beta}\big)
    +c R^{(\mu}_{(\alpha}\delta^{\nu)}_{\beta)}
    +d R\,g_{\alpha\beta}g^{\mu\nu}
    +e R \delta^{\mu\nu}_{\alpha\beta}\,,
\end{equation}
and $a$, $b$, $c$, $d$, and $e$ arbitrary constants. For example, in the de Donder gauge the graviton kinetic operator\footnote{In an isotropic and homogeneous background, the action of this operator on a tensor reduces to the action of the scalar operator on each component of the tensor \cite{Tsamis:2014hra}, suggesting that the cosmological tensorial instability might be removed by dressing the inverse d'Alembertian into its appropriate tensor representation. Strictly speaking, this would take us beyond the starting point action (\ref{eq:NLAction}) [or, otherwise, we would consider the four indices in the representation of the $(1/\Delta)^{\mu\nu}_{\phantom{\mu\nu}\alpha\beta}$ implicitly shuffling those of the $R_{\mu\nu}$]. Explicit construction of such models can be considered as a topic of future study; in this article we focus on the action of a general scalar (derivative) operator $1/\Delta$ on the (Ricci) tensor $R_{\alpha\beta}$.}  
 would correspond to $a=-2$, $b=0$, $c=2$, $d=1/3$, and $e=-4/3$.

In the following, we also allow the differential part of the operator
to assume a more generic form, involving combinations of the curvature invariants and covariant derivatives $\nabla$ that arise in explicit loop computations. We consider simple forms for the functions $g$ and $h$,
\begin{equation}
g(\triangle)\equiv\frac{\bar{M_{1}}^{2}}{6\triangle}\,,\hspace{15mm} h(\triangle)\equiv\frac{\bar{M_{2}}^{2}}{6\triangle}\,,
\end{equation}
with $\bar M_{1}$ and $\bar M_{2}$ mass scales to be determined by observations. These two properties allow us to simplify the action~(\ref{eq:NLAction}) for a Friedman-Lema\^{i}tre-Robertson-Walker (FLRW) background
\begin{equation}\label{eq:FRW}
\dd s^{2}=H^{-2}\dd N^{2}-a^{2}\dd\mathbf{x}^{2}\,,
\end{equation}
where $N\equiv\ln a$ is the number of $e$-folds, $a$ is the scale factor, and $H\equiv\dot a/a$ stands for the Hubble rate with 
the dot denoting derivative with respect to cosmic time. 
Indeed, by noticing that for an FLRW metric in four dimensions the Weyl tensor
\begin{equation}
C_{\mu\nu\alpha\beta}\equiv R_{\mu\nu\alpha\beta}- \left( g_{\mu[ \alpha} 
R_{\beta ] \nu}-g_{\nu [ \alpha} R_{\beta] \mu}\right)+ \frac{1}{3} g_{\mu [ \alpha} g_{\beta ] \nu} R\,
\end{equation}
vanishes, and using the fact that $\triangle$ is by construction metric compatible, we can write
\begin{align}
C_{\mu\nu\alpha\beta}\triangle^{-1}C^{\mu\nu\alpha\beta} =0 \hspace{5mm}\longrightarrow 
\hspace{5mm}\label{eq:riemanntensor}
R_{\mu\nu\alpha\beta}\triangle^{-1}R^{\mu\nu\alpha\beta}  =-\frac{1}{3}R\triangle^{-1}R+
2R_{\alpha\beta}\triangle^{-1}R^{\alpha\beta}\,.
\end{align}
Substituting this relation into Eq.~(\ref{eq:NLAction}) we obtain the simplified action
\begin{equation}
S=\frac{M_{\rm Pl}^2}{2}\int\text{d}^{4}x\sqrt{-g}\left(-R+R F(\triangle)R+
R^{\alpha\beta}g(\triangle)R_{\alpha\beta}\right)+\int\text{d}^{4}x\sqrt{-g}\mathfrak{\mathit{\mathcal{L}_{m}}}\,,\label{eq:NLAction2}
\end{equation}
where we have defined $F(\triangle)\equiv f(\triangle)-\frac{\bar M^2}{18}\triangle^{-1} $\, with $\bar M^2$ being a linear combination of $\bar M^2_{1}$ and $\bar M^2_{2}$. 
For cosmological backgrounds, the Riemann tensor
 does not explicitly contribute to the background evolution\footnote{Note however that it 
contributes at the level of perturbations.}; all the dynamical information can be encoded in nonlocal terms constructed out of Ricci scalars and Ricci tensors only.

The $R F(\triangle)R$ part of Eq.~(\ref{eq:NLAction2}) has been extensively studied the literature for several 
choices of $F(\triangle)$ and $\triangle$ \cite{Cusin:2016nzi,Nersisyan:2016hjh,Deffayet:2009ca,Koivisto:2008xfa,Maggiore:2014sia}.
In this work we concentrate on the phenomenological consequences of the 
tensorial structure $R^{\alpha\beta}g(\triangle)R_{\alpha\beta}$. In particular, we consider the action 
\begin{equation}
S=\frac{M_{\rm Pl}^2}{2}\int\text{d}^{4}x\sqrt{-g}\left(-R+\frac{\bar{M}^{2}}{6}R_{\alpha\beta} \triangle^{-1}
R^{\alpha\beta}\right)
+\int\text{d}^{4}x\sqrt{-g}\mathfrak{\mathit{\mathcal{L}_{m}}}\,,\label{eq:actionten}
\end{equation}
with
\begin{equation}
\triangle\equiv m^{4}+\alpha_{1}\Box+\alpha_{2}\Box^{2}+\beta_{1}R_{\alpha\beta}\nabla^{\alpha}
\nabla^{\beta}+\beta_{2}R\Box+\gamma\left(\nabla^{\alpha}R_{\alpha\beta}\right)\nabla^{\beta}\,,\label{eq:nonlocaloperator}
\end{equation}
and $\alpha_{1}$, $\alpha_{2}$, $\beta_{1}$, $\beta_{2}$, $\gamma$, $m$ constant parameters. Up to the $m^4$ term, the differential operator \eqref{eq:nonlocaloperator} is the most general fourth-order operator containing at least one covariant derivative acting on the function following it.
This choice of operator has a special physical motivation in the celebrated conformal anomaly \cite{Capper:1974ic,Riegert:1984kt}, in which quantum effects break the conformal symmetry of massless fields coupled to gravity. In this case the trace of the energy-momentum tensor receives a nonvanishing contribution from the counterterms introduced by renormalization. The form of this contribution is highly nontrivial and depends on the particle content. In four dimensions, the effective action induced by the conformal anomaly is given by \cite{Riegert:1984kt}
\begin{equation}
S_{\rm A}=-\frac{1}{8}\int \dd^4x\sqrt{-g}\left(E-\frac{2}{3}\Box R\right)\triangle_4^{-1}\left[b' \left(E-\frac{2}{3}\Box R\right)-2b C_{\mu\nu\alpha\beta}^2\right],\label{eq:CAaction}
\end{equation}
where $E\equiv R_{\mu\nu\alpha\beta}^2-4R_{\mu\nu}^2+R^2$ is the Gauss-Bonnet term, $C_{\mu\nu\alpha\beta}^2 = R_{\mu\nu\alpha\beta}^2-2R_{\mu\nu}^2+R^2/3$ is the square of the Weyl tensor, $b$ and $b'$ are numbers that depend on the particle content of the theory, and $\triangle_{4}$ is defined as 
\begin{equation}
\triangle_4=\Box^2+2R_{\alpha\beta}\nabla^{\alpha}\nabla^{\beta}-\frac{2}{3} R\Box +\frac{2}{3}(\nabla^{\alpha}R_{\alpha\beta})\nabla^{\beta}\label{eq:delta4}\,.
\end{equation}
This operator is just a particular case of the operator \eqref{eq:nonlocaloperator} with $m=0$, $\alpha_1=0$, $\alpha_2=1$, $\beta_1=2$, $\beta_2=-2/3$ and $\gamma=2/3$.\footnote{Note that even though the form of the operator \eqref{eq:nonlocaloperator} is motivated by the form of the conformal anomaly operator $\triangle_{4}$, the action~(\ref{eq:NLAction}) considered in this paper is not of the form of the action~(\ref{eq:CAaction}).}\\

The equations of motion associated to the nonlocal action \eqref{eq:actionten} can be obtained by following a standard procedure for the study of nonlocal theories. We localize the action by introducing two auxiliary fields $S_{\alpha\beta}$ and $K_{\alpha\beta}$, defined as solutions of the differential equations
\begin{equation}\label{eq:Nonlocaleq}
\triangle S_{\alpha\beta}  =R_{\alpha\beta}\,,\hspace{10mm}
\Box S_{\alpha\beta} =K_{\alpha\beta}\,.
\end{equation}
After variation of our nonlocal action (\ref{eq:actionten}) with
respect to the metric $g_{\mu\nu}$ and taking into account the identity $\delta\left(\triangle^{-1}\right)=-\triangle^{-1}\delta(\triangle)\triangle^{-1}$ (see \rcite{soussa_nonlocal_2003,Barvinsky:2014lja} for details) we get the modified Einstein equations
\begin{equation}
R_{\alpha\beta}-\frac{1}{2}g_{\alpha\beta}R=\frac{1}{M_{\rm Pl}^{2}}\left(T_{\alpha\beta}+T_{\alpha\beta}^\mathrm{NL}\right)\,,\label{eq:EinsteinHilbert}
\end{equation}
where $T_{\alpha\beta}$ is the energy-momentum tensor associated to the matter Lagrangian ${\cal L}_m$, which is by
construction covariantly conserved,  $\nabla_\alpha T^{\alpha}_{\beta}=0$. The interaction term
$T_{\alpha\beta}^\mathrm{NL}$  arises from the
variation of the nonlocal term $R_{\alpha\beta}\triangle^{-1}R^{\alpha\beta}$ and can be naturally split into six pieces,
\begin{equation}
T_{\alpha\beta}^\mathrm{NL}=T_{\alpha\beta}^{\mathrm{NL}(0)}+T_{\alpha\beta}^{\mathrm{NL}(1)}+T_{\alpha\beta}^{\mathrm{NL}(2)}+T_{\alpha\beta}^{\mathrm{NL}(3)}+T_{\alpha\beta}^{\mathrm{NL}(4)}+T_{\alpha\beta}^{\mathrm{NL}(5)}\label{eq:Nonlocalenergy}\,,
\end{equation}
where we have defined
\begin{align}
\frac{1}{2M^{4}}T_{\alpha\beta}^{\mathrm{NL}(0)} & \equiv\frac{1}{2}R_\mn S^\mn g_{\alpha\beta}-2R_{\alpha}^{\mu}S_{\mu\beta}-\Box S_{\alpha\beta}-g_{\alpha\beta}\nabla_{\mu}\nabla_{\nu}S^{\mn}+2\nabla_{\mu}\nabla_{\alpha}S_{\beta}^{\mu}\,,\\
\frac{1}{2\alpha_{1}M^{4}}T_{\alpha\beta}^{\mathrm{NL}(1)}  &\equiv \frac{1}{2}g_{\alpha\beta}\nabla_{\sigma}S^{\mu\nu}\nabla^{\sigma}S_{\mu\nu} - \nabla_{\alpha}S^{\mu\nu}\nabla_{\beta}S_{\mu\nu}- 2S^{\mu\nu}\nabla_{\nu}\nabla_{\alpha}S_{\mu\beta} + 2 S_{\alpha}^{\mu}\nabla_{\nu}\nabla_{\beta}S_{\mu}^{\nu} \\
&\hphantom{{}=} 
-  2 \nabla_{\mu}S^{\mu\nu}\nabla_{\alpha}S_{\beta\nu} +2\nabla_{\nu}S_{\alpha}^{\mu}\nabla_{\beta}S_{\mu}^{\nu}
+ \frac{1}{2}g_{\alpha\beta}S^{\mu\nu}\nabla_{\sigma}\nabla^{\sigma}S_{\mu\nu}\,,\\
\frac{1}{2\alpha_{2}M^{4}}T_{\alpha\beta}^{\mathrm{NL}(2)}  &\equiv 2K_{\beta\nu}\nabla_{\mu}\nabla_{\alpha}S^{\mu\nu}+2\nabla_{\alpha}S^{\mu\nu}\nabla_{\mu}K_{\beta\nu}-2\nabla_{\mu}S^{\mu\nu}\nabla_{\alpha}K_{\beta\nu}
- 2S^{\mu\nu}\nabla_{\mu}\nabla_{\alpha}K_{\beta\nu}\\
&\hphantom{{}=} -2K^{\mu\nu}\nabla_{\mu}\nabla_{\alpha}S_{\beta\nu}-2\nabla_{\alpha}S_{\beta\nu}\nabla_{\mu}K^{\mu\nu}+  2\nabla_{\mu}S_{\beta\nu}\nabla_{\alpha}K^{\mu\nu}+2S_{\beta\nu}\nabla_{\mu}\nabla_{\alpha}K^{\mu\nu} \nonumber \\
&\hphantom{{}=} -2\nabla_{\alpha}S^{\mu\nu}\nabla_{\beta}K_{\mu\nu}+g_{\alpha\beta}\nabla_{\sigma}S^{\mu\nu}\nabla^{\sigma}K_{\mu\nu}+\frac{1}{2}g_{\alpha\beta}S^{\mu\nu}\Box K_{\mu\nu}+\frac{1}{2}g_{\alpha\beta}K_{\mu\nu}\Box S^{\mu\nu}\,,\nonumber\\
\frac{1}{2\beta_{1}M^{4}}T_{\alpha\beta}^{\mathrm{NL}(3)}  &\equiv-2R_{\alpha\sigma}\nabla_{\mu}S^{\mu\nu}\nabla^{\sigma}S_{\beta\nu}+2R_{\beta\sigma}\nabla_{\mu}S_{\alpha\nu}\nabla^{\sigma}S^{\mu\nu}+2R_{\beta\sigma}S_{\alpha\nu}\nabla_{\mu}\nabla^{\sigma}S^{\mu\nu}+2R_{\alpha\sigma}S^{\mu\nu}\nabla_{\beta}\nabla^{\sigma}S_{\mu\nu}\nonumber\\
 &\hphantom{{}=} + \frac{1}{2}R_{\alpha\beta}\nabla_{\sigma}S^{\mu\nu}\nabla^{\sigma}S_{\mu\nu}+\frac{1}{2}R_{\alpha\beta}S^{\mu\nu}\Box S_{\mu\nu}-\frac{1}{2}\nabla^{\sigma}\nabla_{\alpha}\left(S^{\mu\nu}\nabla_{\beta}\nabla_{\sigma}S_{\mu\nu}\right)- 2R_{\alpha\sigma}S^{\mu\nu}\nabla_{\mu}\nabla^{\sigma}S_{\beta\nu}\nonumber \\
&\hphantom{{}=}-\frac{1}{2}g_{\alpha\beta}\nabla^{\sigma}\nabla^{\tau}\left(S^{\mu\nu}\nabla_{\sigma}\nabla_{\tau}S_{\mu\nu}\right)+S_{\mu\alpha}S_{\beta}^{\mu}\left(\nabla^{\sigma}\nabla^{\tau}R_{\sigma\tau}\right) - \frac{1}{2}\nabla^{\sigma}\nabla_{\beta}\left(S^{\mu\nu}\nabla_{\alpha}\nabla_{\sigma}S_{\mu\nu}\right)\nonumber \\
&\hphantom{{}=} +\frac{1}{2}\Box\left(S^{\mu\nu}\nabla_{\alpha}\nabla_{\beta}S_{\mu\nu}\right) - R_{\alpha\sigma}\nabla_{\beta}S^{\mu\nu}\nabla^{\sigma}S_{\mu\nu} - 2\left(\nabla_{\mu}R^{\mu\sigma}\right)\left(S_{\alpha}^{\nu}\nabla_{\sigma}S_{\beta\nu}\right)\nonumber \\
&\hphantom{{}=}-2\left(\nabla_{\mu}R_{\alpha\sigma}\right)\left(S^{\mu\nu}\nabla_{\sigma}S_{\beta\nu}\right)+2\left(\nabla_{\mu}R_{\beta\sigma}\right)\left(S_{\alpha\nu}\nabla^{\sigma}S^{\mu\nu}\right)\nonumber +2\left(\nabla^{\mu}R_{\mu\sigma}\right)\nabla^{\sigma}\left(S_{\alpha\nu}S_{\beta}^{\nu}\right)\\
&\hphantom{{}=} -\left(\nabla^{\sigma}R_{\sigma\alpha}\right)\left(S_{\mu\nu}\nabla_{\beta}S^{\mu\nu}\right)+\frac{1}{2}\left(\nabla^{\sigma}R_{\alpha\beta}\right)\left(S_{\mu\nu}\nabla_{\sigma}S^{\mu\nu}\right)\,,\\
\frac{1}{2\beta_{2}M^{4}}T_{\alpha\beta}^{\mathrm{NL}(4)}  &\equiv S_{\alpha}^{\nu}\Box RS_{\beta\nu} - RS_{\beta}^{\nu}\Box S_{\alpha\nu}+S_{\beta\nu}\nabla_{\mu}\nabla_{\alpha}RS^{\mu\nu} + \nabla_{\alpha}RS^{\mu\nu}\nabla_{\mu}S_{\beta\nu}-\nabla_{\mu}RS^{\mu\nu}\nabla_{\alpha}S_{\beta\nu} \nonumber \\
&\hphantom{{}=}- S^{\mu\nu}R\nabla_{\mu}\nabla_{\alpha}S_{\beta\nu} - S^{\mu\nu}\nabla_{\mu}\nabla_{\alpha}RS_{\beta\nu}-\nabla_{\alpha}RS_{\beta\nu}\nabla_{\mu}S^{\mu\nu}\nonumber \\
&\hphantom{{}=} + \nabla_{\mu}RS_{\beta\nu}\nabla_{\alpha}S^{\mu\nu} + RS_{\beta\nu}\nabla_{\mu}\nabla_{\alpha}S^{\mu\nu}-\nabla_{\beta}RS^{\mu\nu}\nabla_{\alpha}S_{\mu\nu} + R_{\alpha\beta}\left(S^{\mu\nu}\Box S_{\mu\nu}\right)\nonumber \\
&\hphantom{{}=} + \frac{1}{2}g_{\alpha\beta}\nabla_{\sigma}RS^{\mu\nu}\nabla^{\sigma}S_{\mu\nu} + \frac{1}{2}g_{\alpha\beta}RS^{\mu\nu}\Box S_{\mu\nu} + g_{\alpha\beta}\Box\left(S^{\mu\nu}\Box S_{\mu\nu}\right) - \nabla_{\alpha}\nabla_{\beta}\left(S^{\mu\nu}\Box S_{\mu\nu}\right)\,,\\
\frac{1}{2\gamma M^{4}}  T_{\alpha\beta}^{\mathrm{NL}(5)}&\equiv\frac{1}{2}g_{\alpha\beta}\nabla_{\tau}\left(S^{\mu\nu}R^{\tau\sigma}\nabla_{\sigma}S_{\mu\nu}\right)-\frac{1}{2}\nabla_{\tau}\left(S^{\mu\nu}R_{\alpha\beta}\nabla^{\tau}S_{\mu\nu}\right)\nonumber\\
&\hphantom{{}=}+ S^{\mu\nu}\left(\nabla^{\tau}R_{\tau\alpha}\nabla_{\beta}S_{\mu\nu}\right) + S^{\mu\nu}\left(\nabla_{\beta}R_{\alpha\tau}\nabla^{\tau}S_{\mu\nu}\right)\nonumber \\
&\hphantom{{}=} + \frac{1}{2}\nabla_{\sigma}\nabla_{\alpha}\nabla_{\beta}\left(S^{\mu\nu}\nabla^{\sigma}S_{\mu\nu}\right)-\frac{1}{2}g_{\alpha\beta}\nabla_{\sigma}\nabla_{\tau}\nabla^{\sigma}\left(S^{\mu\nu}\nabla^{\tau}S_{\mu\nu}\right)-\nabla_{\sigma}\left(S_{\alpha\nu}S_{\beta}^{\nu}\left(\nabla_{\mu}R^{\mu\sigma}\right)\right)\,, 
\end{align}
with $M^{4}\equiv\frac{1}{12}\bar{M}^{2} M_{\rm Pl}^{2}$.

\section{ $R_{\alpha\beta}\triangle^{-1}R^{\alpha\beta}$ cosmology}\label{sec:cosmo}

Finding exact solutions for the complicated set of equations derived in the previous section is certainly not an 
easy task.
In what follows, we adopt the approach of \rcite{Ferreira:2013tqn} and assume that the energy density contributed 
by nonlocal effects is subdominant, so that we have the standard radiation- and matter-dominated
eras ($T_{\alpha\beta}^\mathrm{NL}\ll T_{\alpha\beta}$). We investigate the stability of various regions of parameter space, defined as the presence or absence of growing modes in the energy density contributed by the nonlocal interactions.

We assume $\alpha_{1}=m=0$, which allows
us to find certain analytic solutions. We have carried out a preliminary numerical study for nonvanishing values of 
$\alpha_{1}$ and $m$  and found that the inclusion of these parameters does not significantly modify the results presented 
below. A full numerical study of the parameter space is beyond the scope of this work. 

\subsection{Radiation-dominated era}

During radiation domination, the Ricci scalar is $0$ and the terms proportional to $R\,\Box$ and
$\left(\nabla^{\sigma}R_{\sigma\tau}\right)\nabla^{\tau}$ in Eq.~(\ref{eq:nonlocaloperator}) vanish (the latter due to the Bianchi identity). On top of that, the symmetry of the FLRW metric \eqref{eq:FRW} allows us to reduce the tensor $S_{\mu\nu}$ in Eq.~(\ref{eq:Nonlocaleq}) to a simple diagonal form, $S_{\mu}^{\nu}=\mathrm{diag}\left(S_{1},-S_{2},-S_{2},-S_{2}\right)$, that depends on two (homogeneous) scalar functions $S_{1}$ and $S_{2}$. Taking into account these simplifications, the set of equations (\ref{eq:Nonlocaleq}) can be rewritten as
\begin{align}
&\alpha_{2}S_{+}^{(4)}-6\alpha_{2}S_{+}^{(3)}+3\beta_{1}S_{+}''-11\alpha_{2}S_{+}''+(60\alpha_{2}-9\beta_{1})S_{+}'+8(\beta_{1}-4\alpha_{2})S_{+}  =\frac{4a^{4}}{\Omega_{\text{R}}^{0}}\,,\label{eq:nonloccon}\\
&\alpha_{2}S_{-}^{(4)}-6\alpha_{2}S_{-}^{(3)}+(3\beta_{1}+5\alpha_{2})S_{-}''+(12\alpha_{2}-9\beta_{1})S_{-}'  =0\,,\label{eq:nonloccon2}
\end{align}
where $'\equiv \dd/\dd N$ denotes derivatives with respect to the number of $e$-folds $N$, $\Omega_{\text{R}}^{0}$ is the current value of the critical radiation density, and we have defined two dimensionless variables
\begin{equation}\label{dimlessS}
S_{+}\equiv(S_{1}+S_{2})H_{0}^{2}\,,\hspace{10mm} S_{-}\equiv(S_{1}-3S_{2})H_{0}^{2}\,,
\end{equation}
in terms of the Hubble parameter today, $H_{0}^{2}=H^{2}a^{4}/\Omega_{\text{R}}^{0}$. Note that for $\alpha_2=0$, the fourth-order differential 
equations~\eqref{eq:nonloccon} and \eqref{eq:nonloccon2} reduce to second-order differential equations admitting the simple solution
\begin{align}
S_{+} & =a^{\frac{3}{2}}\left[c_{1}\sin\left(\frac{1}{2} \sqrt{\frac{5}{3}}\ln a\right)+c_{2} 
\cos\left(\frac{1}{2} \sqrt{\frac{5}{3}} \ln a\right)\right]+\frac{a^{4}}{5 \beta_{1}\Omega_{\text{R}}^{0}}\,,\label{eq:analyticradalpa0+}\\
S_{-} & =\frac{1}{3}\tilde{c_{1}} a^{3 }+\tilde c_{2}\,,\label{eq:analyticradalpa0-}
\end{align}
where $c_{1}$, $c_2$, $\tilde c_1$, and $\tilde c_2$ are integration constants to be fixed by initial conditions.
In the general case $\alpha_{2}\neq 0$, the solution of Eqs.~(\ref{eq:nonloccon}) and (\ref{eq:nonloccon2}) is
\begin{align}
S_{+} & =a^{3/2}\left(c_{1}a^{-q_{-}}+c_{2}a^{q_{-}}+c_{3}a^{-q_{+}}+c_{4}a^{q_{+}}-\frac{a^{5/2}}{\text{\ensuremath{\Omega_{\text{R}}^{0}}}(24\text{\ensuremath{\alpha_{2}}}-5\text{\ensuremath{\beta_{1}}})}\right)\,,\label{eq:analyticrad}\\
S_{-} & =\frac{2a^{3/2-y/2}}{3-y}\tilde{c_{1}}+\frac{2a^{3/2+y/2}}{3+y}\tilde{c_{2}}+\frac{1}{3}\tilde{c_{3}}a^{3}+\tilde{c_{4}}\,,\label{eq:analyticrads-}
\end{align}
where
\begin{equation}
q_{\mp}  =\frac{\sqrt{49\alpha_{2}-6\beta_{1}\mp2\sqrt{(44\alpha_{2}-9\beta_{1})(12\alpha_{2}-\beta_{1})}}}{2\sqrt{\alpha_{2}}}\,,\label{eq:qpm}\hspace{10mm}
y  =\frac{\sqrt{25\alpha_{2}-12\beta_{1}}}{\sqrt{\alpha_{2}}}\,,
\end{equation}
and $c_{i}$ and $\tilde{c_{i}}$   $(i=1,...,4)$ are integration constants.
Note that in both cases the leading contributions in $S_+$, $S_-$ at large values of the scale factor $a$ take the power-law forms 
\begin{equation}\label{eq:Asimtoticrad}
S_{+}  \thickapprox\tilde{A}a^{A}\,,\hspace{10mm}
S_{-}  \approx\tilde{B}a^{B}\,,
\end{equation}
with $A$ and $B$ being \emph{positive} constants related only to the model parameters $\lbrace\alpha_2,\beta_1\rbrace$, and $\tilde{A}$ and $\tilde{B}$ coefficients 
keeping track of the integration constants $c_{i}$ and $\tilde{c_{i}}$   $(i=1,...,4)$, i.e., keeping track of the initial conditions. Inserting these asymptotic 
expressions into Eq.~(\ref{eq:Nonlocalenergy}) and comparing the result with the
standard form $T^{\mu}_{\nu}=\textrm{diag}(\rho_\mathrm{NL},-p_\mathrm{NL},-p_\mathrm{NL},-p_\mathrm{NL})$ for a perfect fluid, we can derive approximate expressions at the lowest order in $\Omega_{\text{R}}^0$ for the nonlocal energy density $\rho_\mathrm{NL}$ and the nonlocal equation of state $w_\mathrm{NL}\equiv p_\mathrm{NL}/\rho_\mathrm{NL}$ during radiation domination,
\begin{align}
\rho_\mathrm{NL} & \approx -3M^{4}\text{\ensuremath{\Omega_{\text{R}}^{0}}}\left(\tilde{A}(A+4)a^{A-4}+\tilde{B}(B+1)a^{B-4}\right)\,,\label{eq:andennl}\\
w_\mathrm{NL}
& \approx  -\frac{1}{3}\frac{(A-1)\tilde{A}(A+4)a^{A-4}+\tilde{B}\left(B^{2}-1\right)a^{B-4}}{\tilde{A}(A+4)a^{A-4}+\tilde{B}(B+1)a^{B-4}}\,.\label{eq:aneqsnl}\
\end{align}
The behavior of $w_\mathrm{NL}$ at large values of $a$ depends on the relation between $A$ and $B$, i.e., on the precise choice of the model parameters $\lbrace\alpha_2,\beta_1\rbrace$. For $B<A$, the equation 
of state asymptotically approaches $w_\mathrm{NL}=-\frac{1}{3}\left(A-1\right)$, while for $B>A$ it instead evolves towards
$w_\mathrm{NL}=-\frac{1}{3}\left(B-1\right)$. Note that, contrary to the nonlocal energy density $\rho_\mathrm{NL}$,  the 
asymptotic values of $w_\mathrm{NL}$ do not depend on the initial conditions.\\

For $\alpha_{2}=0$ we have $A=4$ and $B=3$ [cf. Eqs.~(\ref{eq:analyticradalpa0+}) and (\ref{eq:analyticradalpa0-})]. These asymptotic values translate into a constant nonlocal energy density $\rho_\mathrm{NL}$ and a cosmological-constantlike equation of state $w_\mathrm{NL}=-1$. Therefore, nonlocal contributions with $\alpha_2=0$  can 
\textit{in principle} lead to a viable cosmology, as long as the radiation energy density is dominant over $\rho_\mathrm{NL}$ for the entire radiation-dominated era.\footnote{Note that this conclusion holds only for $\alpha_{1}=0$. As shown in Ref.~\cite{Ferreira:2013tqn}, the $\alpha_{1}\neq0$ scenario contains growing modes and leads to an unstable cosmology.}

The situation changes completely in the $\alpha_{2}\neq0$ case.
Demanding the absence of a growing mode in Eq.~(\ref{eq:andennl}) imposes $A,B\leq4$.
By considering Eqs.~(\ref{eq:analyticrad}) and (\ref{eq:qpm}) with the restriction $B\leq4$, we get the constraints
\begin{equation}\label{bounds}
\alpha_{2}  >0\,,\hspace{10mm}
\beta_{1}  \in\left[0,\frac{25}{12}\alpha_{2}\right]\,.
\end{equation}
Unfortunately, these two conditions are never satisfied for $A\leq 4$. Indeed, a 
simple inspection of Eq.~(\ref{eq:analyticrad}) shows that in order to keep $A\leq4$ we must 
have $q_{+}+3/2\leq4$ and $q_{-}+3/2\leq4$, or equivalently $\beta_{1}\leq 24/5\,\alpha_{2}$ 
and $\beta_{1}\geq 24/5\,\alpha_{2}$, in clear contradiction with each other and 
with (\ref{bounds}). The growing modes become rapidly dominant unless the prefactor of 
the nonlocal contribution in the action is largely suppressed.\footnote{Note that 
instabilities associated with tensorial structures appear also in ultraviolet extensions of 
general relativity. In the case of Starobinsky inflation, the problem of instabilities coming 
from the tensorial components is addressed by introducing a hierarchy between energy scales of 
the $R^2$ and $R_{\mu\nu}R^{\mu\nu}$ terms \cite{Mukhanov:1981xt}.}\\

\begin{figure}
\centering
\includegraphics[scale=1.1]{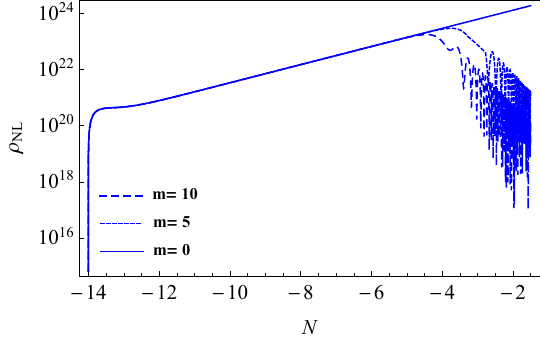}
\caption{Functional behavior of the nonlocal energy density $\rho_\mathrm{NL}$ versus the number of $e$-folds $N$ during radiation domination for an operator $\triangle=\alpha_1\Box+m^4$ and different values of $m$.  All quantities are expressed in units with $H_0=1$. Note that the dimensionful parameter $\alpha_1$ is not an independent parameter: together with $\bar M$, it fixes the amplitude of nonlocal effects and does not modify the dynamics. In this plot, we set $\bar M=H_{0}$ and $\alpha_1=H_{0}^2$. The late-time evolution of the nonlocal energy density develops a damped oscillatory pattern in the vicinity of $N=0$ when our radiation-domination ansatz for the scale factor $a$ is no longer applicable. The \textit{average} of this quantity over an oscillation period scales as $a^{-8}$, i.e., faster than the radiation fluid $(\rho_\text{R}\sim a^{-4})$. A similar damping during radiation domination would require values of $m$ comparable to the Hubble rate during that era.}
 \label{evolrad}
\end{figure}

This conclusion does not seem  to be modified for an operator $\triangle=\alpha_1\Box+m^4$ with values of $m$ and $\alpha_1$ of order $H_0$ and $H_0^2$, respectively. As shown in Fig.~\ref{evolrad}, the evolution of the nonlocal energy density in this case develops a damped oscillatory pattern in the vicinity of $N=0$, when our radiation-domination ansatz for the scale factor $a$ is no longer applicable. A similar damping during radiation domination would require values of $m$ comparable to the Hubble rate at that era.

\subsection{Matter-dominated era}

Can the instabilities generated during radiation domination be suppressed during the subsequent evolution of the Universe? To 
answer this question we study the behavior of a subdominant nonlocal tensorial contribution 
during matter domination ($\rho_{\text{M}}\gg\rho_\mathrm{NL}$). Taking into account the definitions in 
(\ref{dimlessS}) (with $H_{0}^{2}=H^{2}a^{3}/\Omega_{\text{M}}^{0}$), we can write the differential
equations in (\ref{eq:Nonlocaleq}) as
\begin{align}
 & 4\alpha_2 S_{+}^{(4)}-12 \alpha_2S_{+}^{(3)}+6\beta_{1}S_{+}''-73\alpha_2 S_{+}''
 -12\beta_{2}S_{+}'' \label{eq:nonlocconmat+} \\&
-3S_{+}'(9\beta_{1}-6\gamma-41\alpha_{2}+6\beta_{2})+16S_{+}(3\beta_{1}+7\alpha_{2}+
6\beta_{2}) =\frac{12a^{3}}{\Omega_{\text{M}}^{0}}\,,\nonumber \\
&\frac{4}{3}\alpha_2S_{-}^{(4)}-4\alpha_2S_{-}^{(3)}-(3\alpha_{2}-2\beta_{1}+4\beta_{2})S_{-}''
-(9\beta_{1}-6\gamma-9\alpha_{2}+6\beta_{2}) S_{-}'
=-\frac{4a^{3}}{\Omega_{\text{M}}^{0}}\,,\label{eq:nonlocconmat-}
\end{align}
 with  $\Omega_{\text{M}}^{0}$ being the critical matter density today. As in the case of radiation domination, if we choose  $\alpha_2= 0$, then Eqs.~(\ref{eq:nonlocconmat+}) and (\ref{eq:nonlocconmat-}) are reduced from fourth-order to second-order differential equations.
 \begin{figure}
\centering
\includegraphics[scale=1.1]{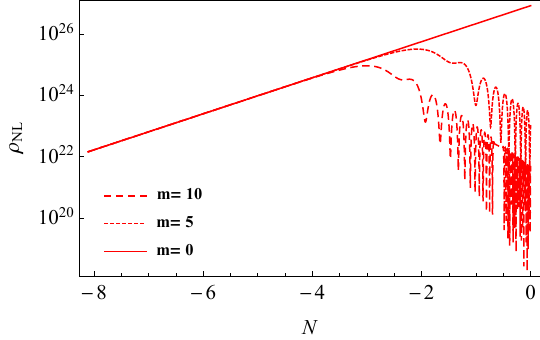}
\caption{Functional behavior of the nonlocal energy density $\rho_\mathrm{NL}$ versus the number of $e$-folds $N$ during matter domination for an operator $\triangle=\alpha_1\Box+m^4$ and different values of $m$.  All quantities are expressed in units with $H_0=1$. Note that the dimensionful parameter $\alpha_1$ is not an independent parameter. Together with $\bar M$ in the action, it fixes the amplitude of nonlocal effects and does not modify the dynamics.  In this plot we set $\bar M=H_{0}$ and $\alpha_1=H_{0}^2$. The late-time evolution of the nonlocal energy density develops a damped oscillatory pattern. The \textit{average} of this quantity over an oscillation period scales as $a^{-6}$,  i.e., faster than the matter fluid $(\rho_M\sim a^{-3})$. Note that when $m$ is of the order of the Hubble rate at matter-radiation equality, this could alleviate the previous growth during radiation domination.}
 \label{evolmat}
\end{figure}
 These equations can be solved analytically,
\begin{align}
S_{+} & =\frac{4a^{3}}{\Omega_{\text{M}}^{0}(7\beta_{1}-22\beta_{2}+18\gamma)}+c_{1}a^{p_{-}}+c_{2}a^{p_{+}}\,,\\
S_{-} & =\frac{4a^{3}}{9\Omega_{\text{M}}^{0}(\beta_{1}+6\beta_{2}-2\gamma)}+\tilde{c_{1}}\frac{2(\beta_{1}-2\beta_{2})}{3(3\beta_{1}+2\beta_{2}-2\gamma)}a^{\tilde{y}}+\tilde{c_{2}}\,,
\end{align}
with
\begin{equation}
p_{\pm}  =\frac{9 \beta_{1}+6\beta_{2}-6 \gamma\pm\sqrt{-47 \beta_{1}^2+108\beta_{1} (\beta_{2}-\gamma )+548 \beta_{2}^{2}-72 \beta_{2}\gamma +36 \gamma ^{2}}}{4 (\beta_{1}-2\beta_{2})}\,,\label{eq:ppm}\hspace{5mm}
\tilde{y}  =\frac{3(3\beta_{1}+2\beta_{2}-2\gamma)}{2(\beta_{1}-2\beta_{2})}\,.
\end{equation}
 The detailed solution of Eqs.~(\ref{eq:nonlocconmat+}) and (\ref{eq:nonlocconmat-}) for the $\alpha_2\neq 0$ case is cumbersome and largely irrelevant for the following discussion. 
 On general grounds, the leading contributions to $S_+$ and $S_-$ at large values of the scale factor $a$ can be
 parametrized as\footnote{Our results cover the tensorial action induced by the conformal anomaly and  the extension 
 of the Maggiore-Mancarella model considered in Ref.~\cite{Cusin:2015rex}. For the parameters 
 associated to the conformal anomaly ($\alpha_1=0$, $\alpha_2=1$, $\beta_1=2$, $\beta_2=-2/3$, $\gamma=2/3$),
one obtains
\begin{align}
S_{+} & =a^{\frac{3}{4}}\left(c_{1}a^{-\frac{1}{4}\sqrt{133-4\sqrt{385}}}+c_{2}a^{\frac{1}{4}\sqrt{133-4\sqrt{385}}}+
c_{3}a^{-\frac{1}{4}\sqrt{133+4\sqrt{385}}}+c_{4}a^{\frac{1}{4}\sqrt{133+4\sqrt{385}}}-
\frac{2a^{9/4}}{9\Omega_{\text{M}}^{0}}\right)\,,\label{eq:analyticmat}\nonumber\\
S_{-} & =2\tilde{c_{1}}a^{\frac{1}{2}}+\frac{2}{3}\tilde{c_{2}}a^{\frac{3}{2}}+\tilde{c_{3}}a
+\tilde{c_{4}}-\frac{2a^{3}}{15\Omega_{\text{M}}^{0}}\,,\nonumber
\end{align}
while for the case $\triangle \propto \Box^{2}$ ($\alpha_1=\beta_1=\beta_2=\gamma=0$, $\alpha_2\neq0$) 
considered in \rcite{Cusin:2015rex}  we find
\begin{align}
S_{+} & =a^{-\frac{1}{4}\left(3+\sqrt{137}\right)}\left(c_{2}a^{\frac{\sqrt{137}}{2}}+c_{3}a^{3}+c_{4}a^{\frac{1}{2}\left(6+\sqrt{137}\right)}+
c_{1}-\frac{3a^{\frac{1}{4}\left(15+\sqrt{137}\right)}}{44\Omega_{\text{M}}^{0}}\right)\,,\nonumber\\
S_{-} & =-\frac{2}{3}\tilde{c_{1}}a^{-\frac{3}{2}}+\frac{2}{3}\tilde{c_{2}}a^{\frac{3}{2}}+\frac{\tilde{c_{3}}}{3}a^{3}+\tilde{c_{4}}-\frac{36\ln a-44}{243\Omega_{\text{M}}^{0}}a^{3}\,.\nonumber
\end{align}}  
\begin{equation}
S_{+} \approx\tilde{C}a^{C}\,,\label{eq:materasymp}\hspace{10mm}
S_{-}  \approx\tilde{D}a^{D}\,,
\end{equation}
with the \textit{positive} constants $C$ and $D$ encoding information about the model parameters, and the 
prefactors $\tilde C$ and $\tilde D$ tracing the initial conditions. Note that the $a^3$ dependence of the 
source term in Eqs.~(\ref{eq:nonlocconmat+}) and (\ref{eq:nonlocconmat-}) forces $C$ and $D$ to be asymptotically  
larger or equal to 3. Using \cref{eq:Nonlocalenergy}, we can derive the nonlocal 
energy density
\begin{equation}
\rho_\mathrm{NL}  \approx\frac{ M^{4}(\Omega_{\text{M}}^{0})^{2}}{32}(\tilde{E}a^{2C-6}+\tilde{F}a^{2D-6})\,,\label{eq:ronlmatter}
\end{equation}
with $\tilde{E}$ and $\tilde{F}$ being some constants built from the free parameters of the 
theory and the initial conditions. Since the exponents $C$ and $D$ satisfy always 
the condition $C,D\geq3$, we have either a constant or growing nonlocal 
energy density $\rho_\mathrm{NL}$.
Therefore, the instabilities arising during radiation domination cannot be suppressed in 
the matter-dominated era. Note that this result also holds for the operator $\triangle=\alpha_1\Box+m^4$ with nonvanishing values of $m$ and $\alpha_1$, with numerical results presented in Fig.~\ref{evolmat}. 

For the sake of completeness, we present in Table~\ref{tab:fixedpoints} the asymptotic values of the nonlocal equation of state $w_\mathrm{NL}$ when only one of the  parameters in the operator \eqref{eq:nonlocaloperator} is different from $0$. The values associated to the conformal anomaly operator  \eqref{eq:delta4} are also displayed. 
This helps us to see in a qualitative way the contribution coming from the different operators in \eqref{eq:nonlocaloperator} when the condition $T_{\alpha\beta}^\mathrm{NL}\ll T_{\alpha\beta}$ is satisfied.\footnote{Note that this can always be achieved by fine-tuning the mass scale $\bar{M}$ in Eq.~(\ref{eq:actionten}).}
\begin{table}[t]
\centering{}%
\begin{tabular}{|c|c|c|}
\hline 
Model & $w_\mathrm{NL}$(RD) &$w_\mathrm{NL}$(MD)\tabularnewline
\hline 
\hline 
$\alpha_{1}$ & $-1.25$ & $-1.45$\tabularnewline
\hline 
$\alpha_{2}$ & $-1.79$ & $-2.45$\tabularnewline
\hline 
$\beta_{1}$ & $-1$ & $-2$\tabularnewline
\hline 
$\beta_{2}$ & $0$ & $-1$\tabularnewline
\hline 
$\gamma$ & $0$ & $-1$\tabularnewline
\hline 
$m$ & $5/3$ & $1$\tabularnewline
\hline 
$\triangle_{4}$ & $-1.55$ & $-1.92$\tabularnewline
\hline 
\end{tabular}\protect\caption{\label{tab:fixedpoints} Characteristic values of the nonlocal equation of state $w_\mathrm{NL}$ during RD and MD when only one of the
 parameters in the nonlocal operator \eqref{eq:nonlocaloperator} is different from $0$. Note that the operators associated to $\beta_2$ and $\gamma$ vanish exactly during radiation domination.
 The values associated to the conformal anomaly operator \eqref{eq:delta4} are also presented.}
\end{table}
\section{Conclusions}\label{sec:conclusions}

In this paper, we have explored the stability of a general class of tensorial nonlocal extensions of
general relativity.  Our result is a direct answer to \rcite{Ferreira:2013tqn}, where the authors  conjectured 
that the instabilities arising in the tensorial $R_{\alpha\beta}\Box^{-1}R^{\alpha\beta}$  model might be cured 
by a generalization of the d'Alembertian operator to $\alpha_1\Box+m^4$ or 
to the conformal anomaly operator $\triangle_4$. 
We have found that the growing mode and the associated instabilities of tensorial nonlocal models cannot be generically avoided by
introducing the most general nonlocal operator at second order in covariant derivatives. 

This conclusion holds also for a restricted version of the operator, namely $\alpha_1\Box+m^4$, if the scale $m$ is chosen to be of the order of the Hubble rate today. One could alternatively consider scenarios in which $m$ is comparable to the Hubble rate at matter-radiation equality. In those cases, an oscillatory pattern arises that could be compatible with our requirement that the nonlocal contribution to the cosmic expansion be subdominant to the matter contribution. This might give rise to phenomenologically interesting features in the form of an oscillating early dark energy. 

In the presence of growing modes, terms at higher and higher order in curvature are expected to become 
relevant, compromising the validity of the effective action \eqref{eq:actionten}. Although one cannot exclude the possibility of some cancellation mechanism among the various terms, a nonperturbative study within the effective nonlocal theory is quite difficult. We believe that the instabilities associated to tensorial nonlocalities should instead be addressed in the framework of local field theories by considering mechanisms able to generate well-behaved nonlocal actions in the infrared. 

\acknowledgments We acknowledge support from DFG through the 
project TRR33 ``The Dark Universe.'' H.N. acknowledges financial support from DAAD through the 
program ``Forschungsstipendium f\"{u}r Doktoranden und Nachwuchswissenschaftler.'' Y.A. acknowledges support from the Netherlands Organization for Scientific Research (NWO) and the Dutch Ministry of Education, Culture and Science (OCW), and also from the D-ITP consortium, a program of the NWO that is funded by the OCW. A.R.S. was supported by funds provided to the Center for Particle Cosmology by the University of Pennsylvania.

\bibliography{dynamicalNLv3Notes,nonlocalRefsTSK,Henrik}

\end{document}